\documentstyle[12pt,epsfig]{article}
\def\be{\begin{equation}} \def\ee{\end{equation}} \def\bea{\begin{eqnarray}}
\def\eea{\end{eqnarray}} \def\nnb{\nonumber}

\begin{document}

\hfill{May 20, 2019,\ \ \ {\tt LCago5}}

\begin{center}
\vskip 6mm 
\noindent
{\Large\bf  
The $S_{E1}$ factor of 
radiative $\alpha$ capture on $^{12}$C 
in cluster effective field theory
}
\vskip 6mm 

\noindent
{\large 
Shung-Ichi Ando\footnote{mailto:sando@sunmoon.ac.kr}, 
\vskip 6mm
\noindent
{\it
School of Mechanical and ICT convergence engineering, \\
Sunmoon University,
Asan, Chungnam 31460,
Republic of Korea
}
}
\end{center}

\vskip 6mm

The $S_{E1}$-factor of radiative $\alpha$-capture on $^{12}$C
is studied in effective field theory up to next-to-leading order.
A modification of the counting rules for the radiative 
capture amplitudes is discussed.
We find that only two unfixed parameters remain in the amplitudes, 
and those two parameters are fitted to the experimental $S_{E1}$ data.
An $S_{E1}$ factor is calculated
at the Gamow-peak energy as $S_{E1}=59\pm 3$~keV$\cdot$b,
and the result is found to be about 30\% smaller than 
the estimates reported recently. 
An uncertainty of the estimate in the present work is also discussed. 

\vskip 5mm 
\noindent PACS(s): 
11.10.Ef, 
24.10.-i, 
25.55.-e, 
26.20.Fj  

\newpage
\vskip 2mm \noindent
{\bf 1. Introduction}

The radiative $\alpha$ capture on carbon-12,
$^{12}C(\alpha,\gamma)^{16}$O, is one of the fundamental
reactions in nuclear-astrophysics, which determines the $C/O$ ratio
created in the stars~\cite{f-rmp84}.
The reaction rate, equivalently the astrophysical $S$-factor, 
of the process at the Gamow peak energy,
$E_G=0.3$~MeV, however, cannot be determined in experiment
due to the Coulomb barrier.
A theoretical model is necessary to employ in order to extrapolate
the reaction rate down to $E_G$ by fitting model parameters
to experimental data typically measured at a few MeV.

In constructing a model for the study, one needs to take account of
excited states of ${}^{16}$O~\cite{bb-npa06},
particularly, two excited bound states for
$l^\pi_{i\mbox{-}th}=1_1^-$ and $2_1^+$
just below the $\alpha$-${}^{12}$C breakup threshold at
$E=-0.045$ and $-0.24$~MeV~\footnote{
The energy $E$ denotes that of the $\alpha$-${}^{12}$C system
in center of mass frame.
}, respectively,
as well as two resonant (second excited) $1_2^-$ and $2_2^+$ states at
$E=2.42$ and $2.68$~MeV, respectively.
The capture reaction to the ground state of ${}^{16}$O
at $E_G$ is expected to be $E1$ and $E2$ transition dominant
due to the subthreshold $1_1^-$ and $2_1^+$ states,
while the resonant $1_2^-$ and $2_2^+$ states play
a dominant role in the available
experimental data at low energies, typically $1\le E \le 3$~MeV.
The main part of the $S$-factor, therefore, 
consists of $S_{E1}$ and $S_{E2}$ from the $E1$ and $E2$ transitions 
along with a small contribution, $S_{casc}$,
from so called cascade transitions.
During a last half century, a lot of experimental and theoretical
studies for the reaction have been carried out.
See Refs.~\cite{bb-npa06,chk-epja15,bk-ppnp16,detal-17} for review.

Theoretical frameworks employed for the study
are mainly categorized into two~\cite{detal-17}: 
the cluster models using generalized coordinate method~\cite{dbh-npa84}
or potential model~\cite{lk-npa85} 
and the phenomenological models using the parameterization 
of Breit-Wigner, $R$-matrix~\cite{lt-rmp58}, 
or $K$-matrix~\cite{hdz-npa76}.
A recent trend of the study is to rely on intensive numerical analysis, 
in which a larger amount of the experimental data relevant 
to the study are accumulated, and a significant number of parameters of 
the models are fitted to the data 
by using computational power~\cite{detal-17,xetal-npa13,aetal-prc15}.
In the present work, to the contrary,
we discuss another approach to estimate the $S$-factor
at $E_G$; we employ a new method for the study
and discuss a calculation of the $S_{E1}$ factor at $E_G$
based on an effective field theory. 

Effective field theories (EFTs)
provide us a model independent and systematic method
for theoretical calculations.
An EFT for a system in question can be built by introducing
a scale which separates relevant degrees of freedom at low energies
from irrelevant ones at high energies.
An effective Lagrangian is written down in terms of the relevant degrees
of freedom and perturbatively expanded
by counting the number of derivatives order by order.
The irrelevant degrees of freedom are integrated out,
and their effect is embedded in coefficients appearing in the Lagrangian.
Thus, a transition amplitude is systematically
calculated by writing down Feynman diagrams,
while the coefficients appearing in the Lagrangian
are fixed by experiment.
For review, one may refer to
Refs.~\cite{bv-arnps02,bh-pr06,m-15,hjp-17}.
For last two decades, various processes essential in nuclear-astrophysics
have been investigated
by constructing EFTs: 
$p(n,\gamma)d$ at BBN energies~\cite{r-npa00,aetal-prc06}
and $pp$ fusion~\cite{kr-npa99,bc-plb01,aetal-plb08,cetal-plb13},
$^3$He($\alpha,\gamma)^7$Be~\cite{hrv-16,znp-18} and
${}^7$Be($p$,$\gamma$)${}^7$B~\cite{znp-prc14,rfhp-epja14}
in the Sun.

In the previous works~\cite{a-epja16,a-prc18,a-18},
we have constructed an EFT of the radiative capture reaction,
$^{12}$C($\alpha$,$\gamma$)$^{16}$O,
derived the counting rules for the reaction at $E_G$,
and fitted some parameters of the theory to
the binding energies of the ground and excited states, 
$0_1^+$, $0_2^+$, $1_1^-$, $2_1^+$, and $3_1^-$ ($l^\pi_{i-th}$) 
states of $^{16}$O
and the phase shift data of the elastic $\alpha$-$^{12}$C scattering 
for $l=0,1,2$, and 3 channels.
(We review and discuss the counting rules
for the radiative capture reaction
in the following sections.)
When fitting the parameters to the phase shift data,
we have introduced resonance energies of $^{16}$O as
a large scale of the theory.
As suggested by Teichmann~\cite{t-pr51},
below the resonance energies, the Breit-Wigner-type parameterization
for resonances can be expanded in powers of the energy, and
one obtains an expression of the elastic scattering amplitude as
the effective range expansion.
In addition,
we have included the effective range parameters up to third order
($n=3$ in powers of $k^{2n}$)
for the $l=0,1,2$ channels and up to fourth order
($n=4$) for the $l=3$ channel because of the modification of the 
counting rules discussed in Ref.~\cite{a-prc18}.
Though the phase shift data below the resonance energies
are reproduced very well by using the fitted parameters,
we find that significant uncertainties
in the elastic scattering amplitudes remain
when interpolating them to $E_G$.

In the present work, we apply the calculation method of EFT to 
the study of the $S_{E1}$-factor of the radiative capture process
up to next-to leading order (NLO).
Inclusion of the photon field into the present formalism is 
straightforward because a photon field abides
in covariant derivatives in the terms of the effective Lagrangian. 
In the standard counting rules, 
we approximately have three structures (momentum dependence)
in the radiative capture amplitudes for $S_{E1}$  
after fitting the effective range 
parameters to the phase shift data of the elastic scattering for $l=1$. 
We discuss a modification of the standard counting rules 
because of an enhancement effect
of the $p$-wave composite $^{16}$O propagator. 
After taking the modification into account,
we have two structures, which are represented by two 
unknown constants, in the radiative capture amplitudes. 
The two constants are fitted to the experimental $S_{E1}$ data, and  
we calculate an $S_{E1}$-factor at $E_G$. 
We find that our result is about 30\% smaller than the other estimates
reported recently, and 
then discuss an uncertainty of the result of the present work 
from higher order terms of the theory.

The present paper is organized as follows: In section 2, the counting
rules of EFT and the effective Lagrangian for the reactions are briefly
reviewed, and the radiative capture amplitudes 
for the initial $p$-wave state and the formula of the $S_{E1}$ factor  
are displayed in section 3. 
In section 4, a modification of the counting rules is 
discussed, and numerical results are presented in section 5.
Finally, in section 6, results and discussion of the present work are
presented.

\vskip 2mm \noindent
{\bf 2. Effective Lagrangian}

In the study of the radiative capture process,
$^{12}$C($\alpha$,$\gamma$)$^{16}$O, at $E_G=0.3$~MeV
employing an EFT, 
one may regard the ground states of $\alpha$ and $^{12}$C
as point-like particles
whereas the first excited states of $\alpha$ and $^{12}$C
are chosen as irrelevant degrees of freedom, from which
a large scale of the theory is determined~\cite{a-epja16}.
Thus the expansion parameter of the theory
is $Q/\Lambda_H \sim 1/3$ where $Q$ denotes a typical
momentum scale $Q\sim k_G$; $k_G$ is the Gamow peak momentum,
$k_G = \sqrt{2\mu E_G}\simeq 41$~MeV, where
$\mu$ is the reduced mass of $\alpha$ and $^{12}$C.
$\Lambda_H$ denotes a large momentum scale
$\Lambda_H\simeq \sqrt{2\mu_4 E_{(4)}}$ or
$\sqrt{2\mu_{12} E_{(12)}}\sim 150$~MeV where
$\mu_4$ is the reduced mass of
one and three-nucleon system and $\mu_{12}$ is that of four and
eight-nucleon system. $E_{(4)}$ and $E_{(12)}$ are
the first excited energies
of $\alpha$ and $^{12}$C, respectively.
By including terms up to next-to-next-to-leading order,
for example, one may have about 10\% theoretical uncertainty
for the amplitude.

The inclusion of the ground state of $^{16}$O, one may think, 
could cause a problem
because the binding energy of $^{16}$O from the $\alpha$ and $^{12}$C 
breakup threshold is 
$B_0=7.162$~MeV, which is larger than the energy of 
the first excited ($2^+_1$) state of $^{12}$C,
$E_{(12)}=4.440$~MeV. 
However, almost all of the energy released through the capture reaction is 
carried away by the outgoing photon, and thus the initial and final 
nuclear states remain in the states at the typical energies. 
The large momentum scale appears in the intermediate state, 
after the photon is emitted, in the $\alpha$-$^{12}$C propagation.
But the binding energy is far below the $\alpha$-$^{12}$C threshold,
thus its physical effect to the $\alpha$-$^{12}$C propagation is small.
In the present work, 
we do not introduce the $^{16}$O ground state 
as a dynamical degree of freedom but
a source field
because the $^{16}$O ground state appears only in the final state.

An effective Lagrangian for the study of the radiative capture
reaction may be written 
as~\cite{a-epja16,bs-npa01,ah-prc05,aetal-prc07,a-epja07}
\bea
{\cal L} &=& \phi_\alpha^\dagger \left(
iD_0 
+\frac{\vec{D}^2}{2m_\alpha}
+ \cdots
\right) \phi_\alpha
+ \phi_C^\dagger\left(
iD_0
+ \frac{\vec{D}^2}{2m_C}
+\cdots
\right)\phi_C
\nnb \\ && +
\sum_{n=0}^3 
C_{n}^{(1)}d_{i}^\dagger 
\left[
iD_0 
+ \frac{\vec{D}^2}{2(m_\alpha+m_C)}
\right]^n d_{i}
- y_{}^{(1)}\left[
d_{i}^\dagger(\phi_\alpha O_i^{(1)} \phi_C)
+ (\phi_\alpha O_i^{(1)} \phi_C)^\dagger d_{i}
\right] 
\nnb \\ && 
- y_{}^{(0)}\left[
\phi_O^\dagger (\phi_\alpha 
\phi_C)
+ (\phi_\alpha 
\phi_C)^\dagger \phi_O 
\right] 
-h^{(1)}\frac{y^{(0)}y^{(1)}}{\mu}\left[
(
{\cal O}_i^{(1)} \phi_O
)^\dagger 
d_i + \mbox{\rm H.c.} 
\right]
+ \cdots
\label{eq;Lagrangian}
\,,
\eea
where $\phi_\alpha$ ($m_\alpha$) and 
$\phi_C$ ($m_C$) are fields (masses) of $\alpha$ and $^{12}$C, 
respectively.  
$\phi_O$ ($\phi_O^\dagger$) is introduced as a source field
for the $^{16}$O ground state in the final (initial) state. 
In the second last term in Eq.~(\ref{eq;Lagrangian}),
for example, $\phi_O$ ($\phi_O^\dagger$) 
destroys (creates) the $^{16}$O ground state, and 
$(\phi_\alpha\phi_C)^\dagger$ [($\phi_\alpha \phi_C$)] fields 
create (destroy) a $s$-wave $\alpha$-$^{12}$C state.
The transition rate between the $^{16}$O ground state and the 
$s$-wave $\alpha$-$^{12}$C state is parameterized 
in terms of the coupling constant $y^{(0)}$, which is fixed 
by using experimental data. 
$D^\mu$ is a covariant derivative,
$D^\mu = \partial^\mu + i {\cal Q}A^\mu$ where
${\cal Q}$ is a charge operator and $A^\mu$ is the photon field.
The dots denote higher order terms. 
$d_i$ is a composite field of $^{16}$O consisting of 
$\alpha$ and $^{12}$C fields for $l=1$ channel.
The operators for the $l=1$ channel are given as 
\bea
&& 
O_{l}^{(1)} = 
i \left(
\frac{\stackrel{\rightarrow}{D}_C}{m_C} -
\frac{\stackrel{\leftarrow}{D}_\alpha}{m_\alpha} 
\right)_i
\,, 
\ \ \ 
{\cal O}_i^{(1)} = \frac{iD_i}{m_O}\,,
\eea 
where $m_O$ is the mass of $^{16}$O in ground state.
The coupling constants, 
$C_n^{(1)}$ with $n=0,1,2$, and 3, are
fixed by using the effective range parameters of elastic $\alpha$-$^{12}$C
scattering for $l=1$, while
the coupling constant $y^{(1)}$ is redundant;
we set it as $y^{(1)}= \sqrt{6\pi\mu}$.\footnote{
In the denominator of the elastic scattering amplitude, 
the couplings appear in the form, $C_n^{(1)}/y^{(1)2}$ with $n=0,1,2,3$, 
and are fitted to the effective range parameters, $1/a_1$, $r_1$, $P_1$,
$Q_1$, for $l=1$, respectively. The $y^{(1)}$ coupling is redundant, and 
one can arbitrarily fix its value.
}
A contact interaction, the $h^{(1)}$ term,
is introduced to renormalize divergence 
from loop diagrams.

\vskip 2mm \noindent
{\bf 3. Amplitudes and the $S_{E1}$ factor}

\begin{figure}
\begin{center}
\includegraphics[width=12cm]{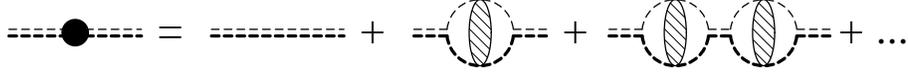}
\caption{
Diagrams for dressed $^{16}$O propagators.
A thick (thin) dashed line represents a propagator of $^{12}$C ($\alpha$),
and a thick and thin double dashed line with and without a filled circle
represent a dressed and bare $^{16}$O propagator, respectively.
A shaded oval represents
a set of diagrams consisting of all possible one-potential-photon-exchange
diagrams up to infinite order and no potential-photon-exchange one.
}
\label{fig;propagator}
\end{center}
\end{figure}

\begin{figure}
\begin{center}
\includegraphics[width=10cm]{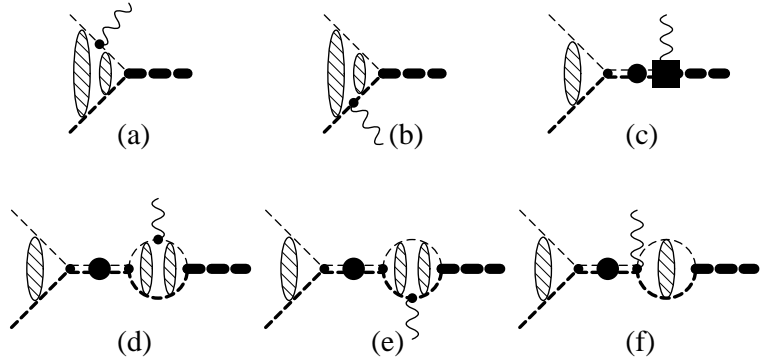}
\caption{
Diagrams for the radiative capture process from the initial $p$-wave 
$\alpha$-$^{12}$C state.
A wavy line denotes the outgoing photon, 
a thick and thin double dashed line with a filled circle in the 
intermediate state,
whose diagrams are displayed in Fig.~\ref{fig;propagator},
the dressed composite $^{16}$O propagator for $l=1$,
and a thick dashed line in the final state 
the ground ($0_1^+$) state of $^{16}$O. 
See the caption of Fig.~\ref{fig;propagator} as well.
}
\label{fig;e1-amplitudes}
\end{center}
\end{figure}

In Fig.~\ref{fig;propagator},
diagrams for dressed composite propagators of $^{16}$O consisting of $\alpha$ 
and $^{12}$C for $l=1$ are depicted, 
in which the Coulomb interaction between
$\alpha$ and $^{12}$C is taken into account~\cite{a-epja16,a-prc18}.
In Fig.~\ref{fig;e1-amplitudes},  diagrams of the radiative 
capture process from the initial $l=1$ state to the $^{16}$O ground ($0_1^+$)
state are depicted,
in which the Coulomb interaction between $\alpha$ and $^{12}$C
is taken into account as well.

The radiative capture amplitude for the initial $l=1$ state is
presented as
\bea
A^{(l=1)} &=& \vec{\epsilon}_{(\gamma)}^* \cdot \hat{p} X^{(l=1)}\,,
\eea
where $\vec{\epsilon}_{(\gamma)}^*$ is the polarization vector 
of outgoing photon and $\hat{p}=\vec{p}/|\vec{p}|$; $\vec{p}$ is
the relative momentum of the initial $\alpha$ and $^{12}$C.
The amplitude $X^{(l=1)}$ can be decomposed as  
\bea
X^{(l=1)} &=& 
X^{(l=1)}_{(a+b)} 
+ X^{(l=1)}_{(c)} 
+ X^{(l=1)}_{(d+e)}
+ X^{(l=1)}_{(f)}\,, 
\eea
where those amplitudes correspond to the diagrams depicted in 
Fig.~\ref{fig;e1-amplitudes}.

We follow the calculation method suggested by 
Ryberg {\it et al.}~\cite{rfhp-prc14},
in which Coulomb Green's functions are represented in coordinate space
satisfying appropriate boundary conditions.
Thus we obtain the expression of those amplitudes in center of mass frame as 
\bea
X_{(a+b)}^{(l=1)} &=& 
2 y^{(0)} 
e^{i\sigma_1} \Gamma(1+\kappa/\gamma_0)
\nnb \\ && \times
\int_0^\infty drrW_{-\kappa/\gamma_0,\frac12}(2\gamma_0 r)
\left[
\frac{Z_\alpha \mu}{m_\alpha}j_0\left(
\frac{\mu}{m_\alpha} k'r
\right)
-\frac{Z_C \mu}{m_C}j_0\left(
\frac{\mu}{m_C} k'r
\right)
\right]
\nnb \\ && \times
\left\{
\frac{\partial }{\partial r}\left[
\frac{F_1(\eta,pr)}{pr}
\right]
+2 \frac{F_1(\eta,pr)}{pr^2}
\right\}\,,
\label{eq;Xab}
\\
X_{(c)}^{(l=1)} &=& +
y^{(0)} 
h^{(1)R}
\frac{6\pi Z_O}{\mu m_O} 
\frac{e^{i\sigma_1}p\sqrt{1+\eta^2}C_\eta}{K_1(p) - 2\kappa H_1(p)}\,,
\label{eq;Xc}
\\
X_{(d+e)}^{(l=1)} &=& 
+i 
\frac{2}{3}y^{(0)}
\frac{
e^{i\sigma_1}p^2\sqrt{1+\eta^2}C_\eta
}{
K_1(p)-2\kappa H_1(p)
}
\Gamma(1+\kappa/\gamma_0)
\Gamma(2+i\eta)
\nnb \\ && \times
\int_{r_C}^\infty drrW_{-\kappa/\gamma_0,\frac12}(2\gamma_0 r)
\left[
\frac{Z_\alpha \mu}{m_\alpha}j_0\left(
\frac{\mu}{m_\alpha} k'r
\right)
-\frac{Z_C \mu}{m_C}j_0\left(
\frac{\mu}{m_C} k'r
\right)
\right]
\nnb \\ && \times
\left\{
\frac{\partial }{\partial r}\left[
\frac{W_{-i\eta,\frac32}(-2ipr)}{r}
\right]
+2\frac{W_{-i\eta,\frac32}(-2ipr)}{r^2}
\right\}\,,
\label{eq;Xde}
\\
X_{(f)}^{(l=1)} &=& -
3 y^{(0)}
\mu
\left[
-2\kappa H(\eta_{b0})
\right]
\left(
\frac{Z_\alpha}{m_\alpha} - \frac{Z_C}{m_C}
\right)
\frac{e^{i\sigma_1}p\sqrt{1+\eta^2}C_\eta}{K_1(p) - 2\kappa H_1(p)}\,,
\label{eq;Xf}
\eea
where $k'$ is the magnitude of outgoing photon momentum, and 
$\kappa$ is the inverse of the Bohr radius, 
$\kappa = Z_\alpha Z_C \mu \alpha_E$, where
$\alpha_E$ is the fine structure constant.
$Z_\alpha$, $Z_C$, and $Z_O$ are the number of protons in 
$\alpha$, $^{12}$C and $^{16}$O, respectively.
$\eta$ is the Sommerfeld parameter, $\eta=\kappa/p$, and 
$\gamma_0$ is the binding momentum of the ground state of $^{16}$O,
$\gamma_0=\sqrt{2\mu B_0}$.
$\Gamma(z)$ and $j_l(x)$ are gamma function and spherical Bessel function,
respectively,
while $F_l(\eta,\rho)$ and $W_{k,\mu}(z)$ are regular Coulomb function
and Whittaker function, respectively.
In addition,
\bea
e^{i\sigma_1} &=& \sqrt{
\frac{\Gamma(2+i\eta)}{
\Gamma(2-i\eta)
}}\,,
\ \ \ 
C_\eta^2 = \frac{2\pi\eta}{e^{2\pi\eta}-1}\,,
\ \ \ 
H_1(p) = p^2(1+\eta^2)H(\eta)\,,
\eea
with
\bea
H(\eta) = \psi(i\eta) + \frac{1}{2i\eta} - \ln(i\eta)\,,
\eea
where $\psi(z)$ is digamma function.

The function $K_1(p)$ contains the information about nuclear interaction
and is represented in terms of the 
effective range parameters of the elastic $\alpha$-$^{12}$C
scattering for $l=1$ as~\footnote{
In the previous work, we had used another parameterization,
so called $v$-parameterization, 
to represent the effective range parameters~\cite{ah-prc12}.
} 
\bea
K_1(p) &=& 
-\frac{1}{a_1} 
+ \frac12 r_1p^2
-\frac14P_1p^4
+Q_1 p^6\,,
\label{eq;K1}
\eea
where $a_1$ is fixed by using the binding energy of the $1_1^-$ 
bound state of $^{16}$O, and other effective range parameters,
$r_1$, $P_1$, and $Q_1$, are fitted to the experimental phase shift data.

Regarding the divergence from the loop diagrams,
the loops of the diagrams (a) and (b) 
in Fig.~\ref{fig;e1-amplitudes} are finite, 
while those of the diagrams (d) and (e) lead to a log divergence 
in $X^{(l=1)}_{(d+e)}$ in the limit, $r\to 0$. 
We introduce a short range cutoff $r_C$
in the $r$ integral in Eq.~(\ref{eq;Xde}), 
and the divergence is renormalized by the counter term,
$h^{(1)}$.
The loop of the diagram (f) diverges and 
is renormalized by the $h^{(1)}$ term as well.
Thus we have 
\bea
h^{(1)R} = h^{(1)} -\mu
\frac{m_O}{Z_O}\left(
\frac{Z_\alpha}{m_\alpha}
-\frac{Z_C}{m_C}
\right)\left[
I_{(d+e)}^{div.}
+ J_0^{div.}
\right]\,,
\eea
where $I_{(d+e)}^{div.}$ is the divergence term from the diagrams (d) and (e)
and $J_0^{div.}$ is that from the diagram (f); we have  
\bea
I_{(d+e)}^{div.} &=& -\frac{\kappa\mu}{9\pi}\int_0^{r_C}\frac{dr}{r}\,,
\ \ \
J_0^{div.} = \frac{\kappa\mu}{2\pi}\left[
\frac{1}{\epsilon}
-3C_E 
+ 2
+\ln\left(
\frac{\pi\mu_{DR}^2}{4\kappa^2}
\right)
\right]\,,
\eea
where, to obtain the expression of $J_0^{div.}$, 
the dimensional regularization in $4-2\epsilon$ space-time dimensions
has been used; $C_E=0.577\cdots$ and $\mu_{DR}$ is a scale factor
from the dimensional regularization.
$h^{(1)R}$ is a renormalized coupling constant 
which is fixed by experiment. 

From the loop diagram (f), 
when the Coulomb interaction is ignored,
the large momentum scale $\gamma_0\simeq 200$~MeV is picked up 
in the numerator of the amplitude. 
It causes the emergence of a term which does not obey the counting 
rules.~\footnote{
A method to renormalize a term which does not obey counting rules 
in an EFT (manifestly Lorentz invariant baryon chiral perturbation theory) 
is known as the extended on mass shell (EOMS) 
scheme~\cite{fgjs-prd03,af-prd07};
one can renormalize the term proportional to $\gamma_0$ in the counter
term, $h^{(1)R}$, even when the Coulomb interaction is ignored.
} 
In the present case, 
the large momentum scale $\gamma_0$ from the ground state energy of $^{16}$O 
appears as a ratio $\kappa/\gamma_0$,
due to the non-perturbative Coulomb interaction, 
where $\kappa$ is another large momentum scale, $\kappa\simeq 245$~MeV. 
The finite term $-2\kappa H(\eta_{b0})$ 
in $X_{(f)}^{(l=1)}$ from the loop of the diagram (f)
where $\eta_{b0}=\kappa/(i\gamma_0)$ 
is reduced to a typical momentum scale, $-2\kappa H(\eta_{b0}) = 25.8$~MeV. 

The $S_{E1}$-factor is defined by
\bea
S_{E1}(E) &=& \sigma_{E1}(E)Ee^{2\pi\eta}\,,
\eea
where the total cross section is 
\bea
\sigma_{E1}(E) &=& \frac43\frac{\alpha_E \mu E_\gamma'}{p
(1+E_\gamma'/m_O)}
|X^{(l=1)}|^2\,, 
\eea
with
\bea
E_\gamma' \simeq B_0 + E - \frac{1}{2m_O}(B_0+E)^2\,.
\eea

\vskip 2mm \noindent
{\bf 4. Modification of the counting rules}

Before fitting the parameters to available experimental data, 
we discuss a modification of the standard counting rules 
for the radiative capture amplitudes. 
An order of an amplitude from each of the diagrams 
is found by counting the number of momenta 
of vertices and propagators in a Feynman diagram.
Thus one has a leading order (LO) amplitude from the diagram (c), 
because the contact $\gamma$-$d_i$-$\phi_O$ vertex of the $h^{(1)R}$ term
does not have a momentum dependence, and 
NLO amplitudes from the other diagrams in Fig.~\ref{fig;e1-amplitudes}.
One may notice a large suppression factor, 
$Z_\alpha/m_\alpha - Z_C/m_C$, appearing in $X^{(l=1)}_{(f)}$;
$(m_O/Z_O)(Z_\alpha/m_\alpha - Z_C/m_C)\simeq -6.5\times 10^{-4}$. 
Similar suppression effect can be found in $X^{(l=1)}_{(a+b)}$ and 
$X^{(l=1)}_{(d+e)}$ as well; we denote those amplitudes as $X^-$, 
and when changing the minus sign to the plus one
in the front of the spherical Bessel function $j_0(z)$
in Eqs.~(\ref{eq;Xab}) and (\ref{eq;Xde}), we do them as $X^+$. 
We thus have 
$|X^{(l=1)-}_{(a+b)}/X^{(l=1)+}_{(a+b)}| \simeq 8.7\times 10^{-4}$ and
$|X^{(l=1)-}_{(d+e)}/X^{(l=1)+}_{(d+e)}| \simeq 3.6\times 10^{-4}$ at the 
energy range, $E=0.9-3$~MeV,
at which we fit the parameters to 
the experimental $S_{E1}$ data in the next section. 
The suppression effect is common among those amplitudes, thus
it does not alter the order counting of the diagrams.

The strong suppression effect mentioned above is well known;
the $E1$ transition is strongly suppressed between isospin-zero
($N=Z$) nuclei. This mechanism is recently reviewed and studied
for $\alpha(d,\gamma)^6$Li reaction by Baye and 
Tursunov~\cite{bt-jpg18}. 
In the standard microscopic calculations with the long-wavelength
approximation, the term proportional to $Z_1/m_1 - Z_2/m_2$ vanishes
because of the standard choice of mass of nuclei as $m_i = A_i m_N$
where $A_i$ is the mass number of $i$-th nuclei and $m_N$ is the 
nucleon mass. We have strongly suppressed but non-zero contribution above
because of the use of the physical masses for $\alpha$ and $^{12}$C.
The small but non-vanishing $E1$ transition for the $N=Z$ cases
has intensively been studied in the microscopic calculations and 
can be accounted by two effects: 
one is the second order term of the 
$E1$ multipole operator in the long-wavelength approximation~\cite{db-plb83}, 
and the other is due to the mixture of the small $T=1$ configuration 
in the actual nuclei~\cite{db-npa86}. 
In the present approach, the first one may be difficult to incorporate
for the point-like particles while the second one could be introduced
from a contribution at high energy: 
At $E\simeq 5$~MeV and 8.5~MeV above the $\alpha$-$^{12}$C breakup threshold, 
$p$-$^{15}$N and $n$-$^{15}$O breakup channels, respectively, are open, and 
$T=1$ resonant states of $^{16}$O start emerging 
(along with the $T=1$ isobars, $^{16}$N, $^{16}$O, and $^{16}$F).
We might have introduced 
the $p$-$^{15}$N and $n$-$^{15}$O fields 
as relevant degrees of freedom in the theory. 
The $p$-$^{15}$N and $n$-$^{15}$O fields, then, appear 
in the intermediate states, 
as $p$-$^{15}$N or $n$-$^{15}$O propagation,
in the loop diagrams (d), (e), (f) 
in Fig.~\ref{fig;e1-amplitudes}
instead of the $\alpha$-$^{12}$C propagation.
One may introduce a mixture of the isospin $T=0$ and $T=1$ states
in the $p$-$^{15}$N or $n$-$^{15}$O propagation,
and the strong $E1$ suppression is circumvented in the loops.
(The contribution from the $p$-$^{15}$N and $n$-$^{15}$O channels 
for the $^{12}$C($\alpha$,$\gamma$)$^{16}$O reaction
has already been studied in the microscopic approach~\cite{db-prc87}.) 
In the present work, however,
the $p$-$^{14}$N and $n$-$^{15}$O fields are regarded 
as irrelevant degrees of freedom at the high energy 
and integrated out of the effective Lagrangian. 
Its effect, thus, is embedded in the coefficient of the contact
interaction, the $h^{(1)R}$ term, in the (c) diagram 
while the $h^{(1)R}$ term is fitted to the experimental $S_{E1}$ data 
in the next section. 

We now discuss an enhancement effect which modifies the counting rules 
for the amplitudes due to the $p$-wave dressed $^{16}$O propagator
in the diagrams (c), (d), (e), and (f). 
The $p$-wave dressed $^{16}$O propagator is enhanced 
due to large cancellations between the effective range terms and terms
generated from the Coulomb self-energy term $H_1(p)$, 
as discussed in Ref.~\cite{a-prc18}. In the standard
counting rules, the order of the propagator 
is assigned to $Q^{-3}$, while because of the inclusion 
up to the $Q_1 p^6$ term in the effective range expansion 
it should be counted as $Q^{-7}$.  
Thus the enhancement factor will be $(Q/\Lambda_H)^{-4} \sim 100$.
To check the magnitude of 
the enhancement effect, we calculate a ratio of the amplitudes and have 
$|(X^{(l=1)}_{(c)}+X^{(l=1)}_{(d+e)}+ X^{(l=1)}_{(f)})/X^{(l=1)}_{(a+b)}| 
\sim 380-30$ at $E=0.1-3$~MeV
after fixing $h^{(1)R}$ to the $S_{E1}$ data.
(Here we have used a value of $h^{(1)R}$ at $r_C=0.1$~fm.)
One can see that the non-pole amplitude $X_{(a+b)}^{(l=1)}$ is significantly
suppressed compared to the other amplitudes
(while we note that the counting rules are applicable at $E_G=0.3$~MeV).

\vskip 2mm \noindent
{\bf 5. Numerical results}

We have five parameters to fit to the data 
in the radiative capture amplitudes; 
three parameters, $r_1$, $P_1$, $Q_1$, are fitted to
phase shift data of the elastic scattering and the other two parameters,
$h^{(1)R}$ and $y^{(0)}$, 
are to the experimental $S_{E1}$ data.
The standard $\chi^2$-fit is performed by employing
a Markov chain Monte Carlo method
for the parameter fitting.~\footnote{
We employ a python package, {\tt emcee}\cite{fmetal-12},
for the fitting.
}
The phase shift data for $l=1$ are taken from 
Tischhauser {\it et al.}'s paper~\cite{tetal-prc09}, 
and the experimental $S_{E1}$ data are 
from the literature
summarized in Tables V and VII in Ref.~\cite{detal-17}:
Dyer and Barnes (1974)~\cite{ex1},
Redder {\it et al.} (1987)~\cite{ex2},
Ouellet {\it et al.} (1996)~\cite{ex3},
Roters {\it et al.} (1999)~\cite{ex4},
Gialanella {\it et al.} (2001)~\cite{ex5},
Kunz {\it et al.} (2001)~\cite{ex6},
Fey (2004)~\cite{ex7},
Makii {\it et al.} (2009)~\cite{ex8},
and Plag {\it et al.} (2012)~\cite{ex9}.

\begin{figure}
\begin{center}
\includegraphics[width=12cm]{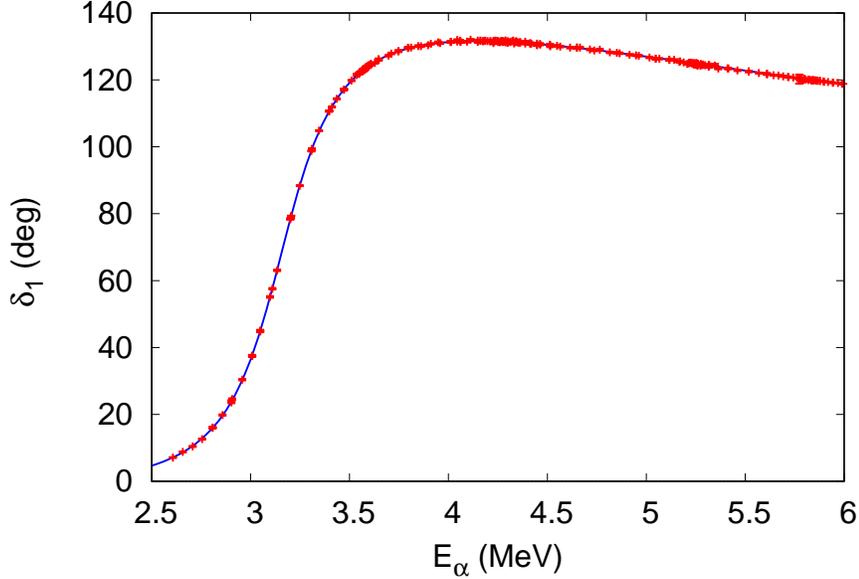}
\caption{
Phase shift, $\delta_1$, plotted by using the fitted effective
range parameters, $r_1$, $P_1$, $Q_1$  as a function of $E_\alpha$. 
The experimental phase shift data are also displayed in the figure.
}
\label{fig;delta1}
\end{center}
\end{figure}

We fit the effective range parameters to the phase shift data
for $l=1$
at $E_\alpha = 2.6-6.0$~MeV and have
\bea
r_1 &=& 0.415272(9)~\mbox{\rm fm$^{-1}$}\,,
\ \ \ 
P_1 = -0.57473(9)~\mbox{\rm fm}\,,
\ \ \
Q_1 = 0.02018(3)~\mbox{\rm fm$^3$}\,,
\eea 
where the number of the data is $N=273$ and 
$\chi^2/N=0.74$. The uncertainties of the fitted values 
stem from those of the experimental data. 
The scattering volume $a_1$ is fixed by using the condition that
the denominator of the elastic scattering amplitude vanishes at the 
binding energy of the $1_1^-$ state, $E=-0.045$~MeV. 
Using the relation in Eq.~(6) in Ref.~\cite{a-prc18} and 
the values of 
the fitted effective range parameters we have
\bea
a_1=-1.658\times 10^5~\mbox{\rm fm$^3$}\,.  
\eea
In Fig.~\ref{fig;delta1}, we plot a curve of the phase shift $\delta_1$
calculated by using the fitted effective range parameters
as a function of $E_\alpha$,
where $E_\alpha$ is the $\alpha$ energy in lab frame~\footnote{
The center of mass energy $E$ is related to $E_\alpha$ by 
$E_\alpha \simeq \frac43E$.
}. 
We display the experimental data in the figure as well.
One can see that the theory curve well reproduces 
the experimental data at the energy range, $E_\alpha=2.6-6.0$~MeV.
Because the energy range is over the energy of 
$1_2^-$ state, $E=2.43$~MeV ($E_\alpha = 3.23$~MeV) and below that of
$1_3^-$ state, $E=5.29$~MeV ($E_\alpha = 7.05$~MeV), our result indicates
that the expression of the effective range expansion given 
in Eq.~(\ref{eq;K1}) is reliable to describe the $1_1^-$ and $1_2^-$ 
states up to $E_\alpha=6.0$~MeV. 

\begin{table}
\begin{center}
\begin{tabular}{c|ccc|c}
$r_C$~(fm) & $h^{(1)R}\times 10^4$~(MeV$^3$) & $y^{(0)}$~(MeV$^{-1/2}$) & 
 $\chi^2/d.o.f$ & $S_{E1}$ (keV b)  \cr \hline
0.01  & $5.2684(11)$ & 0.253(9) & 1.691 & 60.3(18) \cr   
0.035 & $2.4483(11)$ & 0.310(11) & 1.697 & 59.8(18) \cr
0.05 & $1.5294(11)$ & 0.347(12) & 1.700 & 59.3(18) \cr
0.1 & $-0.0695(11)$ & 0.495(18) & 1.715 & 57.9(17) \cr
0.2 & $-1.1909(11)$ & 0.943(34) & 1.763 & 53.6(15) \cr
0.35 & $-1.7106(12)$ & 2.249(84) & 1.926 & 42.1(10) \cr 
\hline 
\end{tabular}
\caption{
Fitted values of the coupling constants, $h^{(1)R}$ and $y^{(0)}$, to
experimental data of $S_{E1}$ at $E=0.9-3$~MeV 
with the cutoff $r_C=0.01-0.35$~fm. 
The number of the data is $N=151$. 
The values in the fourth column are $\chi^2/d.o.f.$ of the fittings,
and those in the last column are our results of $S_{E1}$ at $E_G=0.3$~MeV.
}
\label{table;h1Ry0SE1}
\end{center}
\end{table}

\begin{figure}
\begin{center}
\includegraphics[width=12cm]{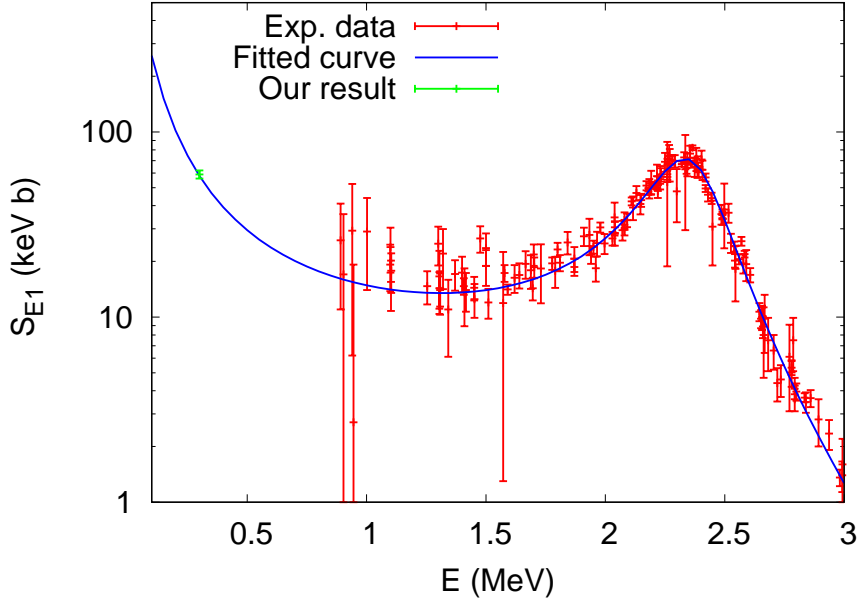}
\caption{
$S_{E1}$ factor plotted by using the fitted parameters 
with $r_C=0.1$~fm as a function of $E$.
The experimental data and our estimate of $S_{E1}$ at $E_G$ 
are also displayed in the figure.
}
\label{fig;se1}
\end{center}
\end{figure}

We fit the parameters, $h^{(1)R}$ and $y^{(0)}$, 
to the experimental data of $S_{E1}$ at the energy range, 
$E=0.9-3.0$~MeV using some values of the cutoff $r_C$
in the range, $r_C=0.01-0.35$~fm,
in the $r$ integral in $X^{(l=1)}_{(d+e)}$ in Eq.~(\ref{eq;Xde}). 
The number of the data is $N=151$.
In Table~\ref{table;h1Ry0SE1}, we display fitted values of 
$h^{(1)R}$ and $y^{(0)}$ along with $\chi^2/d.o.f$ and our estimate
of $S_{E1}$ at $E_G$. 
The uncertainties of the fitted values of $h^{(1)R}$ and $y^{(0)}$ 
stem from those of the experimental data.
We find a significant cutoff dependence of the couplings,
$h^{(1)R}$ and $y^{(0)}$, and the $S_{E1}$ factor at $E_G$
in the table when varying the short range cutoff, $r_C=0.01 - 0.35$~fm; 
as the values of $r_C$ become larger, 
the $\chi^2/d.o.f.$ become larger while the $S_{E1}$ values become smaller.
One may also see that the variation of the $S_{E1}$ factor becomes 
stable at the cutoff values smaller than $r_C=0.1$~fm 
where the $\chi^2/d.o.f$ remain in stable and smaller values.
In Fig.~\ref{fig;se1}, we plot a curve of $S_{E1}$ calculated 
by using the fitted parameters at $r_C=0.1$~fm. 
We display the experimental data and our result of $S_{E1}$
at $E_G$ to be mentioned below in the figure as well. 
One can see that the theory curve reproduces well
the experimental data. 

In the present work, we choose the results of $S_{E1}$ with 
$\chi^2/d.o.f \simeq 1.7$, 
in the cutoff region of the stability of $S_{E1}$, 
for our estimate of $S_{E1}$ 
at the Gamow-peak energy, $E_G=0.3$~MeV, thus, we have
\bea
S_{E1}=59\pm 3\,\  \mbox{\rm keV$\cdot$b}\,,
\eea
where the small, about 5\%, uncertainty 
stems from those of $h^{(1)R}$ and $y^{(0)}$ 
in Table~\ref{table;h1Ry0SE1} as well as that of the $r_C$ dependence 
of $S_{E1}$
within $\chi^2/d.o.f. \simeq 1.7$. 
The previous estimates of the $S_{E1}$ factor at $E_G$ are well summarized 
in Table IV in Ref.~\cite{detal-17}.
The reported vales are scattered from 1 to 340~keV$\cdot$b with various size
of the error bars. 
Nonetheless it is worth pointing out that our result is about 30\% smaller 
than those reported recently: 
$86\pm 22$ by Tang {\it et al.} (2010)~\cite{tetal-prc10}, 
83.4 by Schurmann {\it et al.} (2012)~\cite{setal-12},
$100\pm 28$ by Oulebsir {\it et al} (2012)~\cite{oetal-12},
$80\pm 18$ by Xu {\it et al.} (2013)~\cite{xetal-npa13},
$98.0\pm 7.0$ by An {\it et al.} (2015)~\cite{aetal-prc15}, and
86.3 by deBoer {\it et al.} (2017)~\cite{detal-17}.

Regarding the theoretical uncertainty of the present calculation,
as discussed above, the non-pole amplitude $X_{(a+b)}^{(l=1)}$ 
is suppressed and gives less than one percent correction to $S_{E1}$ 
at $E_G$. 
In the other amplitudes, 
$X_{(c)}^{(l=1)}$, $X_{(d+e)}^{(l=1)}$, and $X_{(f)}^{(l=1)}$,
we find that $X_{(d+e)}^{(l=1)}$ and 
$X_{(c+f)}^{(l=1)} (=X_{(c)}^{(l=1)}+X_{(f)}^{(l=1)}$) have different
momentum dependence and are considerably cancelled with each other.
We have $|(X_{(c+f)}^{(l=1)}+X_{(d+e)}^{(l=1)})/X_{(c+f)}^{(l=1)}|= 
0.055 - 0.023$ at $E = 0.1-3$~MeV; the result is almost linearly 
decreasing as a function of $E=p^2/(2\mu)$. 
It implies that a higher order correction at N$^3$LO effectively 
exists in $X_{(d+e)}^{(l=1)}$ while those two contributions,
$X_{(c+f)}^{(l=1)}$ and $X_{(d+e)}^{l=1)}$, at 
LO+NLO and N$^3$LO equally play a significant role to reproduce
the $S_{E1}$ data. 
Though we have not studied a complete set of the corrections
at N$^3$LO, a next higher order correction
appears at N$^5$LO (because a momentum $\vec{p}$ is vector,
but a correction should be scalar, $p^2$).
Thus the higher order correction at N$^5$LO which we do not have in the
present work may give a few percent correction, 
$(Q/\Lambda_H)^4\simeq 0.012$, to the estimate of $S_{E1}$ at $E_G$.

\vskip 2mm \noindent
{\bf 6. Results and discussion}

In this work, we have applied a framework of EFT to the study
of radiative $\alpha$ capture on $^{12}$C for the first time. 
We have derived the 
radiative capture amplitudes up to NLO in the standard counting rules,
and discussed a modification of the counting rules for the radiative 
capture amplitudes because of the enhancement of the $p$-wave dressed
composite propagator of $^{16}$O. 
We find that the non-pole amplitude $X_{(a+b)}^{(l=1)}$ is significantly
suppressed in the present study.
After taking the modification into account, 
approximately two independent structures (momentum dependence) 
remain in the amplitudes, $X_{(c+f)}^{(l=1)}$ and $X_{(d+e)}^{(l=1)}$,
and we have two unknown parameters in the amplitudes.
The two parameters are fitted to the experimental $S_{E1}$ data at
$E=0.9-3.0$~MeV, and we find the $S_{E1}$ factor, 
$S_{E1}=59\pm 3$~keV$\cdot$b at $E_G$.
Our result is about 30\% smaller than the recent estimates,  
though we have not examined a convergence of our result yet.  

A unique feature of EFT is that one can control a theoretical uncertainty
of a physical quantity in theory. 
In the present work, however, 
we do not examine a convergence of the perturbative expansion of the amplitudes
in the estimate of the $S_{E1}$ factor because we did not include a complete
set of the higher order corrections. 
Thus it is important to study higher order corrections 
to the radiative capture amplitude up to the $Q^4$ order
to check the convergence of the expansion series 
and estimate a theoretical uncertainty of $S_{E1}$ at $E_G$. 
Nonetheless, to accurately fix additional parameters, 
when one includes higher order terms, may not be straightforward 
due to the present quality of the experimental data set of $S_{E1}$.
It might be better studying the other quantities at low energies, 
such as the $\beta$-delayed $\alpha$ emission spectrum of $^{16}$N
or the $\gamma$ angular distribution of the radiative $\alpha$ capture 
process by employing the present EFT approach. 

\vskip 2mm \noindent
{\bf Acknowledgements}

The author would like to thank Kyungsik Kim, 
Young-Ho Song, Youngman Kim, and Renato Higa for useful discussions.
This work was supported by
the Basic Science Research Program through the National Research
Foundation of Korea funded by the Ministry of Education of Korea
(Grant No. NRF-2016R1D1A1B03930122)
and in part by
the National Research Foundation of Korea (NRF)
grant funded by the Korean government
(Grant No. NRF-2013M7A1A1075764 and 
NRF-2016K1A3A7A09005580

\end{document}